\documentclass[aps,pra,twocolumn,nofootinbib]{revtex4-1}
\usepackage{epsfig}
\usepackage[colorlinks,linkcolor=blue,anchorcolor=blue,citecolor=blue,urlcolor=blue,breaklinks=true]{hyperref}
\usepackage{graphics}
\usepackage{color}
\usepackage{amsmath}
\usepackage{bm}
\usepackage{subfig}
\begin{document}
\author{Guo-Hua Liang$^{1}$}
\email{lianggh@njupt.edu.cn}
\author{Ai-Guo Mei$^{1}$}
\author{Zhi-Hui Yang$^{1}$}
\author{Ze-Lin Wei$^{1}$}
\affiliation{$^{1}$ School of Science, Nanjing University of Posts and Telecommunications, Nanjing 210023, China}

\title{Spin textures in curved paths on a curved surface}
\begin{abstract}
This study investigates the quantum dynamics of a spin-1/2 particle confined to a curved path from the dynamics of a two-dimensional curved thin-layer system incorporating spin connection contributions. We demonstrate that the geodesic curvature, normal curvature, and geodesic torsion govern the emergent non-Abelian gauge potential, while the geodesic and Gaussian curvatures govern the effective scalar potential in the Hamiltonian. The resulting spin precession dynamics induced by the gauge potential are analyzed with and without the adiabatic approximation. Under this approximation, the surface topology is linked to the rotation angle of spin orientation along a surface boundary and to the pseudo-magnetic flux. Spin texture evolution along helices illustrates distinct behaviors under geodesic versus non-geodesic propagation. Furthermore, the spin evolution along Viviani's curve exemplifies surface dependence. The curve's topology ensures closure of the spin direction and independence of the spin from the path direction. Our theory establishes a framework for spin-state manipulation via engineered nanostructured channels, enabling novel topological quantum control strategies.
\end{abstract}


\maketitle
\section{INTRODUCTION}
In physics, dimensionality and curvature constitute fundamental concepts that profoundly shape our understanding of natural phenomena. Recent advances in nanostructure synthesis techniques~\cite{meng2021non,Pogosov_2022,lilian2022curvature,Gentile2022,nano12183226} have established versatile platforms for exploring the interplay between dimension reduction and curvature. When the curvature radius becomes comparable to characteristic system lengths, significant modifications emerge in electrical, magnetic, and optical properties. For instance, within the ballistic transport regime, modeling charge carriers in core-shell nanowires as particles confined to curved low-dimensional spaces reveals emergent snake states in band structures~\cite{lauhon2002epitaxial,nl803942p,nl503499w}. Similarly, engineered geometries and curvature in magnetic systems induce nontrivial spin topologies~\cite{PhysRevB.89.180405} and chiral spin states~\cite{0c00720}. Surface waveguides further demonstrate curvature's profound influence, altering optical interference~\cite{RN184}, patterns~\cite{PhysRevLett.105.143901}, phase and group velocities~\cite{Bekenstein2017Control}, and crucially, enabling extreme effective refractive index modifications~\cite{PhysRevResearch.2.013237} unattainable through material composition alone.

Unlike planar systems, quantum particles constrained to curved manifolds exhibit inseparable transverse and tangential dynamics due to broken translational symmetry along the tangential direction. To accurately capture geometric effects in tangential dynamics, the thin-layer procedure~\cite{JENSEN1971586,PhysRevA.23.1982} was developed and has since been applied to diverse systems. This approach yields an effective two-dimensional Schr\"{o}dinger equation incorporating a quantum geometric potential dependent on the surface's intrinsic and extrinsic curvature. Subsequent extensions of this methodology have revealed additional geometric phenomena across multiple domains: Schr\"{o}dinger equations in an electric and magnetic field~\cite{PhysRevLett.100.230403,Brandt_2015,Wang2016a}, Dirac equation~\cite{OUYANG1999297,PhysRevA.48.1861,BRANDT20163036}, non-relativistic spin-1/2 particles~\cite{PhysRevB.91.245412,PhysRevB.64.085330,PhysRevB.87.174413,PhysRevA.90.042117,Wang2017, PhysRevA.98.062112}, Maxwell's equations~\cite{PhysRevA.78.043821,PhysRevA.97.033843,PhysRevA.100.033825}, higher-dimensional systems~\cite{S0217732393000891,Maraner_1995,MARANER1996325,S0217751X97002814,SCHUSTER2003132}, scattering processes~\cite{PhysRevA.107.062806}, magnetic systems~\cite{PhysRevLett.112.257203,0022-3727-49-36-363001} and quantum many-body systems~\cite{PhysRevA.105.023307,SciPostPhysCore.5.1.015,tononi2023low}. Crucially, the inherent degeneracy and internal degrees of freedom in these particle states enable richer geometric dependencies within constrained systems.

While most theoretical studies focus on effective dynamics constrained to 2D curved surfaces or 1D curves embedded in 3D Euclidean space, few address systems embedded within curved manifolds~\cite{rodriguez2024spin}.
Although laboratory creation of curved spacetime remains impractical, lithography~\cite{adfm.202214211}, plasma etching~\cite{huang2024hyperbolic} and laser direct writing~\cite{WANG2021100142} enable experimental realization of 1D structures on curved 2D substrates. In prior work~\cite{LIANG2025170144}, we derived an effective Hamiltonian for a Schr\"{o}dinger particle confined to a curve embedded in a curved surface. Beyond the established geometric potential, this Hamiltonian incorporates a surface geodesic potential, which demonstrates how deviations from the ``straight line" on a curved surface influence constrained quantum dynamics.

This work extends the framework to spin-1/2 particles confined to curved channels on curved thin layers. The inclusion of spin introduces spin connections into the dynamical equations, whose non-Abelian character and interplay with geometries of the surface and curve create rich physical complexity. The investigation also serve as an important expansion on the area of spintronics with nanoscale curved geometries~\cite{gentile2022electronic}.

The outline of the paper is as follows. In Sec.~\ref{sec2}, we present the derivation of the effective Hamiltonian for a spin-1/2 particle constrained to a curve from a curved thin-layer. In Sec.~\ref{sec3} we investigate the spin precession in general situation and demonstrate the spin orientation evolution for two cases: 1) geodesic and non-geodesic lines; 2) the same curve on different curved surfaces. In Sec.~\ref{sec4} we present our conclusions.

\section{Effective Hamiltonian}\label{sec2}
\begin{figure}
  \centering
  \includegraphics[width=0.45\textwidth]{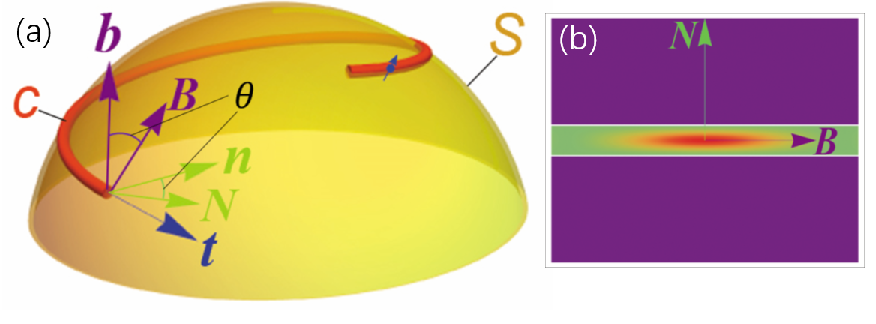}\\
  \caption{(a) Illustration of a curve $\mathcal{C}$ lies on a curved surface $\mathcal{S}$. At a point on $\mathcal{C}$, $\bm{t}$, $\bm{n}$ and $\bm{b}$ are the unit tangential, normal and binormal vector, respectively. $\bm{N}$ is the unit normal vector of $\mathcal{S}$ at this point. There is an angle $\theta$ such that a rotation in the plane of $\bm{n}$ and $\bm{b}$ produces the pair $\bm{N}$ and $\bm{B}$. (b) Schematic of the quantum state probability density in the plane of $\bm{N}$ and $\bm{B}$ for a particle confined to $\mathcal{C}$ embedded in $\mathcal{S}$. }\label{fig1}
\end{figure}

In this section, we consider a spin-1/2 particle constrained to a curve $\mathcal{C}$ embedded in a curved surface $\mathcal{S}$ (see Fig.~\ref{fig1}(a)). Within 3D Euclidean space, we parametrize $\mathcal{C}$ using the Frenet frame $(\bm{t}(s),\bm{n}(s),\bm{b}(s))$, where $\bm{t}$, $\bm{n}$ and $\bm{b}$ denote the tangential, normal and binormal vectors of the curve, and $s$ represents the arclength of $\mathcal{C}$. These vectors satisfy the Frenet-Serret equation
\begin{equation}\label{fse}
\left(
\begin{array}{ccc}
\dot{\textbf{t}}\\
\dot{\textbf{n}}\\
\dot{\textbf{b}}
\end{array}
\right)=\left(
\begin{array}{ccc}
0&\kappa(s)&0\\
-\kappa(s)&0&\tau(s)\\
0&-\tau(s)&0
\end{array}
\right)\left(
\begin{array}{ccc}
\textbf{t}\\
\textbf{n}\\
\textbf{b}
\end{array}
\right),
\end{equation}
where $\kappa(s)$ and $\tau(s)$ are the curvature and torsion of the curve.

At each point on $C$, there exists a normal vector $\bm{N}$ to the surface $\mathcal{S}$. If $\mathcal{C}$ is not a geodesic, the angle $\theta $ between $\bm{n}$ and $\bm{N}$ is non-zero. Since $\mathcal{C}$ lies on $\mathcal{S}$, $\bm{N}$ must lies in the normal plane spanned by $\bm{n}$ and $\bm{b}$. Rotating the Frenet frame by angle $\theta $ within this plane causes $\bm{n}$ to coincide with $\bm{N}$, while $\bm{b}$ transforms into $\bm{B}$ (a vector tangent to $\mathcal{S}$). Then we get the Darboux frame, namely
\begin{equation}\label{tnb}
\left(
\begin{array}{ccc}
{\bm{t}}\\
{\bm{N}}\\
{\bm{B}}
\end{array}
\right)=\left(
\begin{array}{ccc}
1&0&0\\
0&\cos \theta &-\sin \theta \\
0&\sin\theta&\cos\theta
\end{array}
\right)\left(
\begin{array}{ccc}
\bm{t}\\
\bm{n}\\
\bm{b}
\end{array}
\right).
\end{equation}
From Eq.~\eqref{fse}, the Darboux frame satisfies
\begin{equation}\label{dse}
\left(
\begin{array}{ccc}
\dot{\bm{t}}\\
\dot{\bm{N}}\\
\dot{\bm{B}}
\end{array}
\right)=\left(
\begin{array}{ccc}
0&\kappa_n(s)&\kappa_g(s) \\
-\kappa_n(s)&0&-\tau_g(s)\\
-\kappa_g(s)&\tau_g(s)&0
\end{array}
\right)\left(
\begin{array}{ccc}
\bm{t}\\
\bm{N}\\
\bm{B}
\end{array}
\right),
\end{equation}
where $\kappa_g=\kappa\sin\theta$ denotes the geodesic curvature (quantifying curve bending within the surface), $\kappa_n=\kappa\cos \theta$ is the normal curvature, and $\tau_g=\tau-\partial_s\theta$ represents the geodesic torsion (characterizing the curve's twist relative to the surface tangent plane).

Now we construct a Fermi coordinate system to describe a point $P$ on the surface near $\mathcal{C}$. Consider a geodesic emitted from the point $\mathcal{C}(s)$ on the curve in the direction of $\textbf{B}(s)$, which reaches $P$ after propagating a distance $q$. Point $P$ is then mapped to the tangent plane at $\mathcal{C}(s)$ via the exponential map~\cite{klingenberg2013course},
\begin{equation}\label{exp}
P(s,q)=\exp_{\mathcal{C}(s)}[q\textbf{B}(s)].
\end{equation}
In this coordinate system, the metric tensor is given by
\begin{equation}
g_{ab}=\left[
\begin{array}{ccc}
(1-\kappa_g q)^2-K(s,0)q^2+O(q^3) & 0  \\
0 & 1
\end{array}
\right],
\end{equation}
where $K(s,q)$ denotes the Gaussian curvature of the surface.
A detailed derivation of this metric is provided in the appendix~\ref{app}.
Utilizing these geometric descriptions, we proceed to derive the effective Hamiltonian for a spin-1/2 particle constrained to $\mathcal{C}$ from the dynamics on $\mathcal{S}$.

The scenario we consider is that a quantum particle is initially confined to a curved surface $\mathcal{S}$ by a strong normal potential $V_s$, which enforces adiabatic separation: permitting tangential motion while maintaining the ground state along the normal direction $\bm{N}$. A weaker in-plane confining potential $V_c(q)$ further restricts motion to a designated curve $\mathcal{C}$ on $\mathcal{S}$ within the studied energy regime. The relative strength $V_c \ll V_s$ produces an elliptically elongated probability density along the $\bm{B}$-direction in the normal plane of the curve (see Fig.~\ref{fig1}(b)). Such in-plane confinement may originate from thickness variations in curved-layer protrusions~\cite{LIANG2025170144} or engineered material defects along $\mathcal{C}$.

We start from the effective 2D Hamiltonian for a spin-1/2 particle constrained to a curved surface~\cite{PhysRevA.107.022213,Liang_2024}, which has the form
\begin{equation}\label{h2d}
H_{2D}=-\frac{\hbar^2}{2m}\frac{1}{\sqrt{|g|}}D_a(\sqrt{|g|}g^{ab}D_b)+V_g(s,q),
\end{equation}
where $a,b=1,2$ denote 2D curved space indices, the geometric potential $V_g=-\frac{\hbar^2}{2m}(M^2-K)$ with $M$ being the mean curvature of the surface,
and $D_a=\partial_a+\bar{\Omega}_a$, wherein the connection
\begin{equation}
\bar{\Omega}_a=\Omega_a+i(A_{so})_a,
\end{equation}
with
\begin{equation}
\begin{aligned}
\Omega_a=\frac{i}{4}\epsilon^{ij}\omega_{aij}\sigma_N \\
A_{so}=\frac{1}{2\sqrt{g}}\epsilon^{bc}\sigma_b \alpha_{ac}.
\end{aligned}
\end{equation}
Here, $\alpha_{ac}$ is the Weingarten curvature tensor of the surface, $\sigma_N=\bm{\sigma}\cdot \bm{N}$ with $\bm{\sigma}$ the usual Pauli matrices $[\sigma_x,\sigma_y,\sigma_z]$, and the spin connection term
\begin{equation}
\omega_{aij}=e_{\ i}^b(\partial_a e_{bj}-\Gamma_{ab}^c e_{cj}),
\end{equation}
where $e_{\ i}^b$ are zweibeins (components of the frame field) with $i,j=1,2$ denote the local flat space indices, and $\Gamma_{ab}^c$ are the Christoffel symbols. Zweibeins satisfy the relation $g_{ab}=e_{a}^{\ i}e_b^{\ j}\delta_{ij}$ and $e_{a}^{\ i} e_{\ j}^a=\delta_j^i$.

The Hamiltonian in Eq.~\eqref{h2d} is derived from the non-relativistic dynamics of a spin-1/2 particle in 3D Euclidean space using the thin-layer procedure. Due to confinement along the surface normal direction, the tangential connection separates into two distinct components: $\Omega_a$ and $i(A_{so})_a$. Here, $\Omega_a$ governs SU(2) rotations about the surface normal, while $i(A_{so})_a$ couples to rotations within the tangent plane.

Note that the frame field is dependent on the choice of the local flat space coordinates, which is similar to a gauge choice and does not affect the physical result of the system.
Based on the metric, in the curvilinear coordinates $(s,q)$, we can simply choose
\begin{equation}
e_a^{\ i}=\left[
\begin{array}{ccc}
\sqrt{(1-\kappa_g q)^2 -K q^2} & 0 \\
0 & 1
\end{array}
\right]
\end{equation}
and
\begin{equation}
e_{\ i}^a=\left[
\begin{array}{ccc}
\frac{1}{\sqrt{(1-\kappa_g q)^2 -K q^2} } & 0 \\
0 & 1
\end{array}
\right].
\end{equation}
Also, we need give the explicit form of the Christoffel symbols
\begin{equation}
\begin{aligned}
\Gamma_{ss}^{s}=&-q\frac{\partial_s\kappa_g}{g}+\frac{q^2}{2g}(2\kappa_g\partial_s\kappa_g- \partial_s K) , \\
\Gamma_{ss}^{q}=&\kappa_g-q(\kappa_g^2-K), \\
\Gamma_{sq}^{s}=&\Gamma_{qs}^{s}=-\frac{\kappa_g}{g}+q\frac{\kappa_g^2-K}{g}, \\
\Gamma_{qq}^{a}=&\Gamma_{sq}^q=\Gamma_{qs}^q=0.
\end{aligned}
\end{equation}

The spin connection terms are anti-symmetric in its internal indices, then we need only calculate
\begin{equation}
\begin{aligned}
\omega_{s12}=&-\omega_{s21}=\frac{1}{\sqrt{g}}[-\kappa_g+q(\kappa_g^2-K)] \\
\omega_{q12}=&-\omega_{q21}=0.
\end{aligned}
\end{equation}
From the quantities above, we are able to obtain
\begin{equation}
\Omega_s=\frac{i}{2\sqrt{g}}[-\kappa_g+q(\kappa_g^2-K)]\sigma_N,
\end{equation}
and
\begin{equation}
\Omega_q=0.
\end{equation}
Therefore,
\begin{equation}
\begin{aligned}
\bar{\Omega}_s=&\frac{i}{2\sqrt{g}}[-\kappa_g+q(\kappa_g^2-K)]\sigma_N +\frac{i}{2\sqrt{g}}\epsilon^{bc}\sigma_b \alpha_{sc}, \\
\bar{\Omega}_q=&\frac{i}{2\sqrt{g}}\epsilon^{bc}\sigma_b \alpha_{qc}.
\end{aligned}
\end{equation}

In the next step we exert the confining potential $V_c(q)$ to squeeze the 2D motion on the surface to a 1D curve. To perform this task, $V_c(q)$ should have a deep minimum at $q=0$, leading to the concentration of the dynamics in the vicinity of $q=0$.

The Hamiltonian with the confining potential becomes
\begin{equation}
\begin{aligned}
H=&-\frac{\hbar^2}{2m}\frac{1}{\sqrt{g}}[(\partial_s+\bar{\Omega}_s)\sqrt{g}g^{ss}(\partial_s+\bar{\Omega}_s)+ \\ &(\partial_q+\bar{\Omega}_q) \sqrt{g}(\partial_q+\bar{\Omega}_q)]+V_g(s,q)+V_c(q).
\end{aligned}
\end{equation}
To perform the thin-layer procedure, we need find out the the limit of $q\rightarrow0$ for the dynamic equation $H\Psi=E\Psi$.
We assume the Weingarten curvature matrix can be expand as a power series in $q$ about $q=0$, which means in the vicinity of the curve,
\begin{equation}
\alpha_{ab}(s,q)=\sum_{n=0}^{\infty} q^n \frac{\alpha_{ab}^{(n)}(s,0)}{n!}
\end{equation}
where $\alpha_{ab}^{(n)}(s,0)=\partial_q^n \alpha_{ab}(s,q)|_{q=0}$.

Accordingly, the gauge terms can be expanded as
\begin{equation}
\bar{\Omega}_s=\bar{\Omega}_s^{(0)}+ q \bar{\Omega}_s^{(1)}+O(q^2),
\end{equation}
and
\begin{equation}
\bar{\Omega}_q=\bar{\Omega}_q^{(0)}+ q \bar{\Omega}_q^{(1)}+O(q^2),
\end{equation}
Note that in differential geometry, the normal curvature of the curve $\kappa_n=-\alpha_{ss}^{(0)}$ and the geodesic torsion $\tau_g=-\alpha_{sq}^{(0)}$. Then we find
\begin{equation}\label{oms}
\bar{\Omega}_s^{(0)}=\frac{i}{2}(-\kappa_g \sigma_N +\kappa_n\sigma_q-\tau_g \sigma_s).
\end{equation}
Here, $\sigma_q=\bm{\sigma}\cdot \bm{B}$ and $\sigma_s=\bm{\sigma}\cdot \bm{t}$. In the spirit of the thin-layer procedure, we introduce a small dimensionless parameter $\epsilon$ to characterize the energy scales.
Due to quantum state confinement near $q=0$, we rescale the coordinate $q\rightarrow \epsilon q$. Consequently, the confining potential must scale as $V_c \rightarrow V_c \epsilon^{-2}$ to match the energy scale of normal-direction excitations. Typically, the potential $V_c$ can be approximated as a one-dimensional harmonic oscillator potential with its minimum at $q=0$. The wavefunction $\Psi$ satisfies the normalization $\int |\Psi|^2 \sqrt{|g|}dsdq=1$. To obtain an effective 1D wavefunction, we define $\psi=|g|^{1/4}\Psi$ which fulfills $\int |\psi|^2 ds dq=1$. Here, $\int |\psi|^2 dq$ represents the probability density along the curve. The corresponding Hamiltonian for $\psi$ becomes $g^{\frac{1}{4}}Hg^{-\frac{1}{4}}$, which can be expanded in powers of $\epsilon$
\begin{equation}
\begin{aligned}
g^{\frac{1}{4}}Hg^{-\frac{1}{4}}= \frac{1}{\epsilon^2}H_{(-2)}+\frac{1}{\epsilon}H_{(-1)}+H_{(0)}+O(\epsilon)
\end{aligned}
\end{equation}
where
\begin{equation}
H_{(-2)}=-\frac{\hbar^2}{2m}\partial_q^2+V_c,
\end{equation}
\begin{equation}
H_{(-1)}=-\frac{\hbar^2}{m}\bar{\Omega}_q^{(0)}\partial_q,
\end{equation}
and
\begin{equation}\label{h0}
H_{(0)}=-\frac{\hbar^2}{2m}(\partial_s+\bar{\Omega}_s^{(0)})^2-\frac{\hbar^2}{2m}[(\partial_q \bar{\Omega}_q)+q\bar{\Omega}_q^{(1)}\partial_q]+
V_{g}+V_{sg},
\end{equation}
with the the surface geodesic potential
\begin{equation}
V_{sg}=-\frac{\hbar^2}{8m}(\kappa_g^2+2K).
\end{equation}
Through this expansion, we expect to separate the dynamics into tangential and transverse components relative to the curve.
Obviously, $H_{(-2)}$ is just a 1D Hamiltonian for a particle confined by $V_c(q)$. In the limit $\epsilon\rightarrow0$, the transverse state of $H_{(-2)}$ remains in its ground state within the considered energy scale. While the tangential component $H_{(0)}$ depends solely on $s$, the presence of $H_{(-1)}$ prevents dynamical separation due to its q-dependence. To resolve this, we introduce an appropriate gauge transformation for $H$ as follows.

In SU(2) gauge theory, under an infinitesimal gauge transformation parameterized by an infinitesimal vector $\gamma_b$, the fermion field transform as $\psi\rightarrow (1+\gamma_b \sigma^b/2)\psi$, while the gauge field transforms as $\bar{\Omega}_a \rightarrow \bar{\Omega}_a+\partial_a(\gamma_b \sigma^b/2)+i[\gamma_b \sigma^b/2,\bar{\Omega}_a]$. 
Here, for the gauge field $\bar{\Omega}_q$, we define the parameter as
\begin{equation}\label{gau}
\gamma_b \frac{\sigma^b}{2}=- q\frac{i}{\sqrt{g}}\epsilon^{bc}\alpha_{qc}\frac{\sigma_b}{2}+\frac{1}{2}q^2 \partial_q (\frac{i}{2\sqrt{g}}\epsilon^{bc}\alpha_{qc}\sigma_b).
\end{equation}
Applying this gauge transformation, we find
\begin{equation}\label{boq}
\bar{\Omega}_q \rightarrow O(\epsilon^2)
\end{equation}
and
\begin{equation}\label{bos}
\bar{\Omega}_s \rightarrow \bar{\Omega}_s+O(\epsilon).
\end{equation}
The details of obtaining Eq.~\eqref{boq} are put in the Appendix \ref{appb}. To safely obtain Eq.~\eqref{bos}, we have to impose an additional condition: the rate of curvature change along the studied curve must not be too rapid, satisfying $\partial_s \alpha_{qc}\ll \alpha_{qc}/\epsilon$.

The gauge transformation results in $H_{(-1)}\rightarrow O(\epsilon)$ and
\begin{equation}
H_{(0)}\rightarrow H_{\text{eff}}=-\frac{\hbar^2}{2m}(\partial_s+\bar{\Omega}_s^{(0)})^2+
V_{g}+V_{sg}.
\end{equation}
Here, the second term in Eq.~\eqref{h0} becomes first-order in $\epsilon$ after gauge transformation and is therefore omitted from $H_{\text{eff}}$.
At this point, we eventually derive the effective Hamiltonian $H_{\text{eff}}$ for a spin-1/2 particle confined to a curve based on the motion dynamics on a curved surface. The geometric potential $V_g=V_g(s,q=0)$ is determined by the surface curvature at the corresponding point on the curve, while the surface geodesic potential $V_{sg}$ depends on the geodesic curvature and geodesic torsion of the curve. these two potentials emerge in the system of scalar particles as well. When it comes to a spin-1/2 particle, a distinctive feature is the appearance of the gauge term $\bar{\Omega}_s^{(0)}$ in the effective Hamiltonian, which exerts a significant influence on the evolution of spin orientation. This particular gauge field is closely related to the geodesic curvature and the surface curvature matrix. It is important to note that, in this scenario, the geometry of the surface or the curve alone is insufficient to fully determine the effective dynamics of the particle.
\section{Spin precession and geometric phase}\label{sec3}
In this section, we investigate the spatial evolution of the local spin orientation along particular curves on curved surfaces. Because the gauge potential in the effective Hamiltonian $H_{\text{eff}}$ is non-Abelian, the SU(2) phase factor accumulated along a curve from point $s=s_1$ to $s=s_2$ is given by the path-ordered integral $U=\mathcal{P}e^{-\int_{s_1}^{s_2} \bar{\Omega}_s^{(0)} ds} $, where $\mathcal{P}$ denotes path ordering. Considering the form of $\bar{\Omega}_s^{(0)}$, we can  express the gauge potential in vector form as
\begin{equation}
U=\mathcal{P}e^{\frac{i}{2}\int \bm{\sigma}\cdot\bm{\beta} ds}
\end{equation}
where $\bm{\beta}=(\tau_g,\kappa_g,-\kappa_n)$ can be treated as an effective spin-orbit field in the Darboux frame.
For problems with a fixed path, the non-Abelian gauge transformation $\psi_t=U\psi_s$ yields a simplified one-dimensional Schr\"{o}dinger equation
\begin{equation}
-\frac{\hbar^2}{2m}\partial_s^2 \psi_s +(V_g+V_{sg})\psi_s=E\psi_s.
\end{equation}
This shows that spin precession is fully described by the operator $U$.


In general, $\bm{\beta}$ is $s$-dependent, indicating non-commutativity $[\bm{\sigma}\cdot\bm{\beta}(s), \bm{\sigma}\cdot\bm{\beta}(s')]\neq 0$. For arbitrary $\beta(s)$, spin precession must be calculated through numerical discretization
\begin{equation}\label{upr}
U\approx \prod_{j=1}^{n}e^{\frac{i}{2} \bm{\sigma}\cdot\bm{\beta}(j\Delta s) \Delta s},
\end{equation}
where $\Delta s=s/n$ is the step size, and $n$ is the number of segments. For each sufficiently small interval $[s_j,s_{j+1}]$ with $s_j=j \Delta s$, $\bm{\beta}(s_j)$ and $\bm{\beta}(s_{j+1})$ are nearly parallel. This allows us to use the Abelian-like form
\begin{equation}\label{uni}
U_j=e^{\frac{i}{2} \bm{\sigma}\cdot\bm{\beta}(s_j) \Delta s}=\cos \frac{\Delta\Phi_j}{2}+i\bm{\sigma}\cdot \bar{\bm{\beta}}_j\sin\frac{\Delta\Phi_j}{2} ,
\end{equation}
where $\Delta \Phi_j=\sqrt{\kappa^2(s_j)+\tau_g^2(s_j)}\Delta s$ and $\bar{\bm{\beta}}_j=\bm{\beta}(s_j)/\Delta \Phi_j$ is the unit vector in the direction of $\bm{\beta}(s_j)$.

If the direction of $\bm{\beta}$ varies slowly, satisfying $|\partial_s \bar{\bm{\beta}}(s)|\ll 1$, we can employ the adiabatic approximation, where non-commutative terms are neglected. Within this approximation, the operator can be expressed as
\begin{equation}\label{uni}
U_{\text{ad}}=e^{\frac{i}{2}\bm{\sigma}\cdot \bm{h} \Phi}=\cos \frac{\Phi}{2}+i\bm{\sigma}\cdot \bm{h} \sin\frac{\Phi}{2} ,
\end{equation}
where $\Phi=\sqrt{\phi_N^2+\phi_q^2+\phi_s^2}$ with $\phi_N=\int \kappa_g ds$, $\phi_q=-\int \kappa_n ds$, $\phi_s=\int \tau_g ds$, and $\bm{h}=(\phi_s,\phi_N,\phi_q)/\Phi$ is a unit vector in the Darboux frame, which can be regarded as the instantaneous axis of rotation. By using the operator $U$, we can calculate the spin orientation
\begin{equation}
\langle \bm{\sigma} \rangle= \langle \psi_s|U^\dagger \bm{\sigma} U|\psi_s\rangle
\end{equation}
along various curves on different curved surfaces.

It is interesting to give the spatial derivative of the expectation value of the spin components
\begin{equation}\label{pss}
\partial_s\langle \bm{\sigma}\rangle=\langle [\bar{\Omega}_s^{(0)},\bm{\sigma}]\rangle +\langle \partial_s \bm{\sigma} \rangle=2\langle \partial_s \bm{\sigma} \rangle=2\bm{\beta}\times \langle\bm{\sigma}\rangle,
\end{equation}
This general result shows that the rotation rate of spin orientation along the curve is twice of that of the Darboux frame in Eq.\eqref{dse}.

According to the geometric phase theory in degenerate systems~\cite{PhysRevLett.52.2111}, any cyclic adiabatic evolution in parameter space $\lambda^\mu$ generates a Wilczek-Zee phase
\begin{equation}
U=\mathcal{P}\oint A_\mu d\lambda^\mu,
\end{equation}
where $A_\mu$ is a non-Abelian Berry connection arising from state overlaps in the degenerate subspace. In the confinement system we consider, when projected into real space,  Eq.~\eqref{uni} corresponds to the Wilczek-Zee phase factor, with $-\bar{\Omega}_s^{(0)}$ acting as the non-Abelian Berry connection. For a closed curve $\mathcal{C}$, the trace $\text{tr}(U)$ defines a gauge invariant quantity known as Wilson loop. This trace quantifies how surface and curve geometry imprint on quantum spin states, corresponding to measurable interference effects. In the adiabatic approximation, we can find
\begin{equation}\label{wil}
\text{tr}(U_{\text{ad}})=2\cos \frac{\Phi}{2}.
\end{equation}

It is also interesting to note that the Gauss-Bonnet theorem in differential geometry is expressed as
\begin{equation}
\int_M K dA+\int_{\partial M}\kappa_g ds=2\pi \chi(M),
\end{equation}
where $M$ is a compact 2D Riemannian manifold with boundary $\partial M$ and $\chi(M)$ denotes its the Euler characteristic of $M$. This fundamental formula links the curvature of a general surface to its underlying topology. In curved 2D systems, the curvature-induced pseudo-magnetic field $\mathcal{B}$ is proportional to the Gaussian curvature $K$~\cite{PhysRevA.98.062112}, specifically $\mathcal{B}=\hbar K/(2e)$. From Eq.~\eqref{uni}, in the adiabatic approximation, the integral of the geodesic curvature along $\partial M$ corresponds to the rotation angle $\phi_N$, representing the projection of the total angle $\Phi$ onto the plane normal to $\bm{N}$: $\phi_N=\Phi (\bm{h}\cdot \bm{N})$. Consequently, the pseudo-magnetic flux through surface $M$ is
\begin{equation}\label{gbf}
\int_M \mathcal{B} dA= \left[\chi(M)-\frac{\phi_N(\partial_M)}{2\pi}\right]\Phi_0,
\end{equation}
where $\Phi_0=h/(2e)$ is the magnetic flux quantum and $\phi_N(\partial_M)$ is the rotation angle of the particle's spin orientation about the axis $\bm{N}$ as it traverses a complete circuit along the boundary $\partial_M$. This expression connects the pseudo-magnetic field to both surface topology and boundary spin rotation. For any closed curve on a sphere, the integrated Gaussian curvature over the enclosed region ranges continuously between 0 to $2\pi$. Thus, via Eq.~\eqref{gbf}, we can engineer spherical curves to precisely control spin orientation.

In the following analysis, we investigate two illustrative cases:: 1) Spin precession along helical paths on a cylindrical surface, demonstrating distinct geometric effects between geodesic and non-geodesic trajectories; 2) Viviani's curve, exhibiting spin precession behavior when embedded on different surfaces. We further examine path-order dependence and spin interference phenomena.

\subsection{Geodesic and non-geodesic lines: Helices on a cylindrical surface}
We consider three helices on a cylindrical surface with increasing, constant, and decreasing pitch profiles, respectively. Their parameter descriptions are
\begin{equation}
\begin{aligned}
\mathcal{C}_1 : x=&\rho \cos\phi ,y=\rho \sin \phi, z=c\phi; \\
\mathcal{C}_2 : x=&\rho \cos\phi, y=\rho \sin \phi, z=c f [\exp(\phi/f)-1]; \\
\mathcal{C}_3 : x=&\rho \cos\phi, y=\rho \sin\phi, z=c f \ln (\phi/f+1),
\end{aligned}
\end{equation}
where $\rho$ and $\phi$ are the radius and polar angle of the cylindrical surface,respectively. Parameters $c$ and $f$ control the pitch variation rate. The geometric potential $V_g=-\frac{\hbar^2}{8m\rho^2}$ for the three curves are the same since it is determined by the surface geometry. Through stagewise calculations, we derive $\bar{\Omega}_s^{(0)}$, $\bm{\beta}$,  and $V_{sg}$ for these curves,
\begin{equation}
\begin{aligned}
\mathcal{C}_1 : \bar{\Omega}_s&=\frac{i}{2}\left( -\frac{\rho}{\rho^2+c^2}\sigma_q+ \frac{c}{\rho^2+c^2}\sigma_s\right), \\ \bm{\beta}&=\left(-\frac{c}{\rho^2+c^2},0,\frac{\rho}{\rho^2+c^2}\right),  \ \ V_{sg}=0;
\end{aligned}
\end{equation}
\begin{equation}
\begin{aligned}
\mathcal{C}_2 : &\bar{\Omega}_s=\frac{i}{2t_2^2}\left(-\frac{c e^{\frac{\phi}{f}}\rho }{ft_2}\sigma_N -\rho\sigma_q +c e^{\frac{\phi}{f}} \sigma_s \right), \\
&\bm{\beta}=\frac{1}{t_2^2} \left(-c e^{\phi/f},\frac{c e^{\frac{\phi}{f}}\rho }{t_2}, \rho \right), \\
&V_{sg}=-\frac{\hbar^2}{8m}\frac{c^2 e^{2\phi/f}\rho^2}{f^2(c^2 e^{2\phi/f}+\rho^2)^{3}};
\end{aligned}
\end{equation}
and
\begin{equation}
\begin{aligned}
\mathcal{C}_3 :&\bar{\Omega}_s=\frac{i}{2t_3^2} \left(-\frac{cf\rho}{t_3(f+\phi)^2}\sigma_N
 -\rho\sigma_q+ \frac{fc}{\phi+f}\sigma_s \right), \\
&\bm{\beta}=\frac{1}{t_3^2} \left( -\frac{fc}{\phi+f},\frac{cf\rho}{t_3(f+\phi)^2},\rho \right),\\
&V_{sg}=-\frac{\hbar^2}{8m}\frac{c^2 f^2\rho^2 (f + \phi)^2 }{(c^2 f^2 + (f + \phi)^2 \rho^2)^{3}}.
\end{aligned}
\end{equation}
where $t_2=\sqrt{\rho^2+c^2 e^{\frac{2\phi}{f}}}$ and $t_3=\sqrt{\rho^2+\frac{c^2}{(\frac{\phi}{f}+1)^2}}$.

We focus on spin orientation rotation, which is unaffected by scalar potentials. Hence, we assume energy $E\gg |V_g+V_{sg}|$, indicating that $\psi_s$ can be approximated as a plane wave. In the following, we choose the parameters to be $c=\rho=1$ and $f=5$. From the geometric quantities above, one readily notes that $|\partial_s \bar{\bm{\beta}}|\ll 1$ holds for all three curves ( $|\partial_s \bar{\bm{\beta}}|=0$ for $\mathcal{C}_1$). Therefore, we calculate the spin orientation evolution along the three channels on the cylindrical layer by utilizing Eq.~\eqref{uni} in the adiabatic approximation.

\begin{figure}
  \centering
  \includegraphics[width=0.45\textwidth]{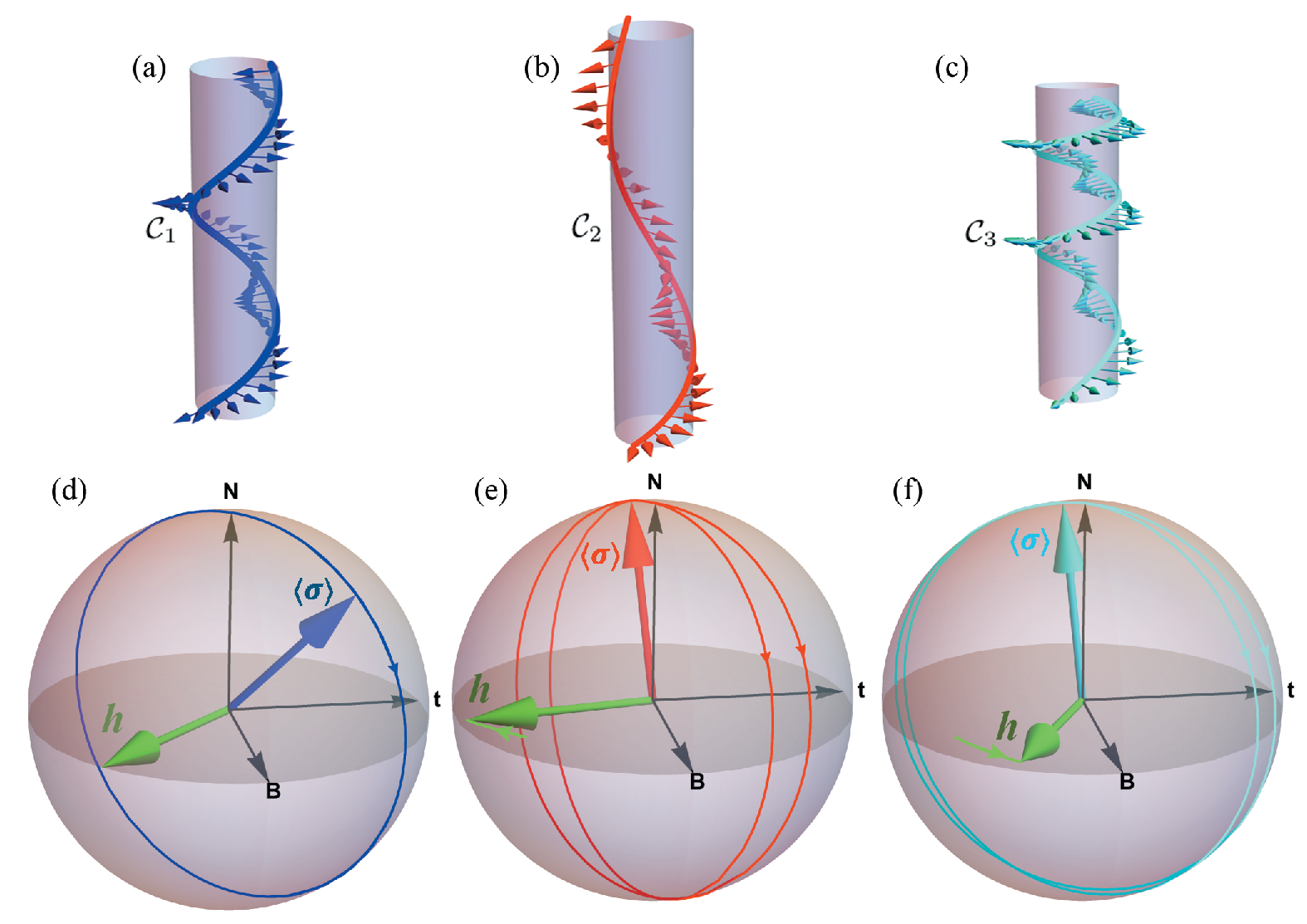}
  \caption{(a)-(c) Evolution of the 3D spin textures along $\mathcal{C}_1$, $\mathcal{C}_2$ and $\mathcal{C}_3$ embedded in a cylindrical surface, respectively. (d)-(f) Evolution of the spin orientation trajectories on the Bloch sphere for $\mathcal{C}_1$, $\mathcal{C}_2$ and $\mathcal{C}_3$, respectively. The initial spin orientation is in $\bm{N}$ direction. Parameters are $c=\rho=1$ and $f=5$.}\label{fig2}
\end{figure}

As shown in Fig.~\ref{fig2}, spin texture rotation along the three helices exhibits distinct rates. $\mathcal{C}_1$ as a geodesic line on the cylindrical surface, has zero geodesic curvature. This constrains the rotation axis $\bm{h}$ to the $\textbf{t}$-$\textbf{B}$ plane, with the effective spin-orbit field $\bar{\bm{\beta}}=\bm{h}=-\bm{e}_z$. In contrast, $\mathcal{C}_2$ and $\mathcal{C}_3$ are non-geodesic curves whose instantaneous rotation axes lie outside this plane. Due to their pitch variations, the $\bm{h}$-axes for $\mathcal{C}_2$ and $\mathcal{C}_3$ are $s$-dependent and evolve in opposite directions, causing the trajectories in Fig.~\ref{fig2}(e) and (f) to skew differentially. Further calculations reveal that greater deviation from the geodesic (smaller $f$)accelerates the spin orientation's departure from circular trajectories in Fig.~\ref{fig2}(d).

Fig.~\ref{fig2}(a) shows that particle motion from $\phi=0$ to $\phi=\pi$ rotates the spin orientation by $2\pi$. This is a manifestation of Eq.~\eqref{pss}, which can also be understood from the view of the total angular momentum conservation. For the curve $\mathcal{C}_1$, the orbital angular momentum component $L_z$ is conserved while $L_x$ and $L_y$ are spatial dependent. One can verify that the rotation axis of the vector $ [\langle L_x \rangle,\langle L_y \rangle ]$ is $\bm{e}_z$, consistent with $\partial_s\langle \bm{J} \rangle=\partial_s(\langle \bm{L} \rangle+\langle \bm{\sigma}/2 \rangle)=0$.

\subsection{Same curve on different surfaces: Viviani's curve}
\begin{figure}
  \centering
  \includegraphics[width=0.45\textwidth]{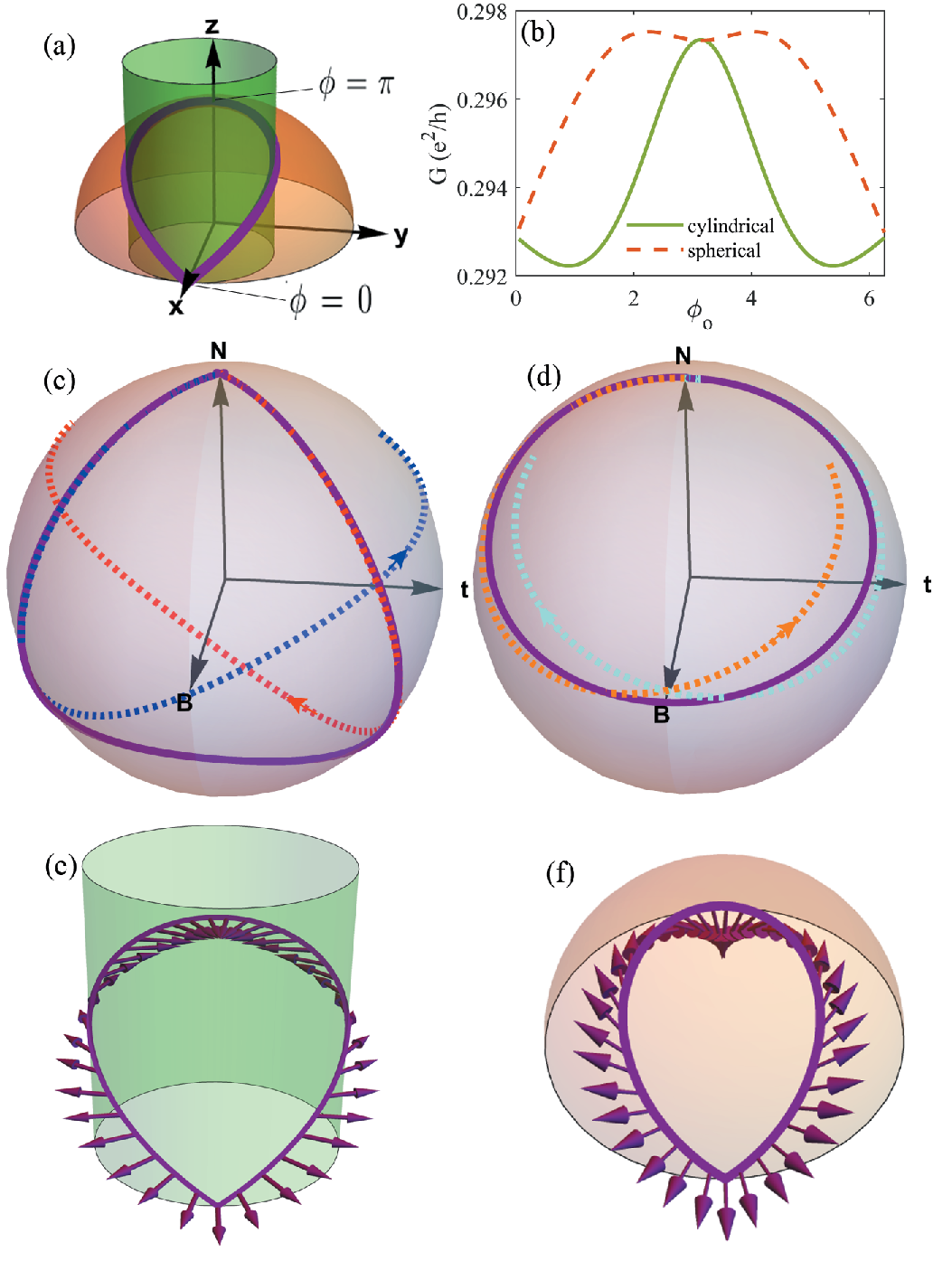}
  \caption{(a) Viviani's curve as the intersecting line of a cylindrical surface and a spherical surface.(b) Conductance profiles for the closed Viviani's curve on cylindrical (solid line) and spherical (dashed line) surfaces as functions of output position $\phi_o$, with input position fixed at $\phi_i=0$. (c) and (d) show the evolution of the spin orientation trajectories on the Bloch sphere for Viviani's curve on the cylindrical surface and spherical surface, respectively. The solid lines and the dashed lines are calculated by numerical discretization and the adiabatic approximation formula, respectively. The solid lines representing evolution in clockwise and counterclockwise directions coincide. (e) and (f) show the evolution of the 3D spin textures along Viviani's curve clockwise and counterclockwise on the cylindrical surface and spherical surface, respectively.  }\label{fig3}
\end{figure}

In this subsection, we discuss spin texture rotation along a curve embedded in different surfaces. When embedded in distinct surfaces, the same curve exhibits differing geodesic curvature, normal curvature, and geodesic torsion, resulting in different spin precession behaviors. Here, the curve is chosen to be Viviani's curve, which is the intersecting line of a cylinder with radius $\rho=1$ and a sphere with radius $r=2$ (see Fig.~\ref{fig3}(a)). The line can be parameterized as
\begin{equation}
\bm{r}=(1+\cos \phi)\bm{e}_x+\sin \phi \bm{e}_y+2\sin\frac{\phi}{2} \bm{e}_z,
\end{equation}
where $\phi$ is the polar angle of the cylinder.

Through calculations on geometry, we can find $\bar{\Omega}_s^{(0)}$, $\bm{\beta}$, $V_{sg}$ and $V_g$ for the curve on different surfaces are
\begin{equation}
\begin{aligned}
&\bar{\Omega}_{s(c)}^{(0)}=\frac{i}{2}\frac{1}{3+\cos \phi}\left(\frac{\sqrt{2}\sin \frac{\phi}{2}}{\sqrt{3+\cos \phi}}
\sigma_N -2\sigma_q+2\cos \frac{\phi}{2} \sigma_s \right), \\
&\bm{\beta}_{(c)}=\frac{1}{3+\cos \phi}\left(-2\cos \frac{\phi}{2}, -\frac{\sqrt{2}\sin \frac{\phi}{2}}{\sqrt{3+\cos \phi}}, 2 \right), \\
&V_{sg(c)}=-\frac{\hbar^2}{4m}\frac{\sin^2 \frac{\phi}{2}}{(3+\cos\phi)^3}, \ \ V_{g(c)}=-\frac{\hbar^2}{8m};
\end{aligned}
\end{equation}
and
\begin{equation}
\begin{aligned}
&\bar{\Omega}_{s(s)}^{(0)}=\frac{i}{2}\left(-\frac{9\sin \frac{\phi}{2}+\sin \frac{3\phi}{2}}{2\sqrt{2}(3+\cos \phi)^{3/2}}\sigma_N-\frac{1}{2}\sigma_q \right),\\
&\bm{\beta}_{(s)}=\left(0,\frac{9\sin \frac{\phi}{2}+\sin \frac{3\phi}{2}}{2\sqrt{2}(3+\cos \phi)^{3/2}},\frac{1}{2} \right),\\
&V_{sg(s)}=-\frac{\hbar^2}{8m}\frac{13+3\cos \phi}{(3+\cos\phi)^3}, \ \ V_{g(s)}=0,
\end{aligned}
\end{equation} 
where the index $(c)$ and $(s)$ stand for the geometric quantities on the cylindrical surface and spherical surface, respectively.

Starting from identical initial spin states (x-axis oriented) at $\phi=0$, Fig.~\ref{fig3}(c) and (d) show the evolution of spin orientation trajectories on the Bloch sphere for Viviani's curve on cylindrical and spherical surfaces, respectively. Solid lines represent calculations using Eq.~\eqref{upr}, while dashed lines correspond to the adiabatic approximation in Eq.~\eqref{uni}. In both figures, solid and dashed trajectories initially coincide but progressively diverge beyond a certain propagation distance. This deviation demonstrates the effect of path-ordered integration of non-commutative terms in $U$, indicating breakdown of the adiabatic approximation in this system.

More importantly, for both clockwise ($\phi=2\pi \rightarrow 0$) and counterclockwise ($\phi=0\rightarrow 2\pi$) propagation, solid trajectories coincide to form a closed shape, whereas dashed lines remain distinct and fail to close in either case. Due to Viviani's curve inherent geometric symmetry, spin orientation trajectories exhibit $\bm{N}$-$\bm{B}$ plane symmetry. Paradoxically, despite the path-ordering dependence of accumulated SU(2) phase factors along curved paths, spin evolution shows no propagation-direction dependence. This may be speculatively attributed to the constraints imposed by the geometric parameters of closed curves on the spin connection over the curved surface. Introducing additional spin-orbit interaction immediately eliminates both direction-independence and closure. Note that using the adiabatic approximation while neglecting non-commutative terms yields fundamentally different results.

Comparing solid lines in Fig.~\ref{fig3}(c) and (d), reveals surface-dependent evolution pathways: confinements on cylindrical and spherical surfaces produce fundamentally different spin dynamics. This is also clear when we comparing the evolution of the 3D spin textures along the same curve in Fig.~\ref{fig3}(e) and (f). This confirms that spin evolution along curves is governed by the synergistic geometry of both the host surface and the constrained path.

Treating the closed curve as an interferometer with fixed input position $\phi_i=0$, we vary output position $\phi_o$ from $0$ to $2\pi$ to quantify geometric control of spin interference. Our analysis confirms that spin orientation at $\phi_o$ matches that of one-sided propagation, demonstrating propagation-direction independence for closed curves. We examine conductance $G=\frac{e^2}{h}\text{Tr}(\bm{T}^\dagger \bm{T})$ via the Landauer formula~\cite{datta1997electronic}, where $\bm{T}$ is the transmission matrix for $|\uparrow \rangle $ and $|\downarrow \rangle$. Fig.~\ref{fig3} (b) plots conductance versus $\phi_o$ for cylindrical and spherical surfaces. It can be observed that the conductances indeed varies with $\phi_o$, but with a very slight amplitude. This is due to the system's effective spin-orbit field is $\phi$-dependent but its strength does not change with $\phi_o$. The solid-dashed distinction reveals different $\phi$-dependence of the effective spin-orbit interaction for the cylindrical and spherical systems.

\section{Conclusion}\label{sec4}
We have unveiled the effective dynamics of spin-1/2 particles constrained to curves within curved thin-layer nanostructures. By extending the thin-layer approach in the Darboux frame, we derived a novel 1D Hamiltonian featuring an SU(2) gauge potential and scalar effective potential. Crucially, the SU(2) gauge field is governed by the curve's geodesic curvature, normal curvature, and geodesic torsion, while the scalar potential encodes surface curvature and curve geometry. Our analysis further deciphers spin precession dynamics, revealing a geometric phase in real space. Strikingly, we discovered that for curved surfaces, the adiabatic boundary spin rotation angle around the normal axis links pseudo-magnetic flux to topology. The spin precession manifests vividly in two paradigmatic systems: 1) Spin evolution along cylindrical helices (contrasting geodesic vs. non-geodesic paths) and 2) Viviani's curve embedded in both cylindrical and spherical surfaces.
It shows that the instantaneous spin rotation axis lies within the $\bm{t}$-$\bm{B}$ plane for geodesics but escapes it for non-geodesics. The same curve generates distinct spin textures when embedded on different curved surfaces, and when the curve is closed, its spin orientation exhibits no dependence on propagation direction.
Collectively, our findings establish a fundamental correspondence between emergent spin textures and quantum transport in low-dimensional nanostructures-bridging geometry, topology, and spintronics.

\acknowledgments

This work is supported in part by the National Natural Science Foundation of China (under Grants No. 12104239), the Natural Science Foundation of Jiangsu Province (under Grant No. BK20210581)

\appendix
\section{Detailed derivation of the 2D metric tensor expansion in Fermi coordinates}\label{app}
Using the exponential map $\exp_{\mathcal{C}(s)}:T_{\mathcal{C}(s)}(\mathcal{S})\rightarrow \mathcal{S}, \textbf{B}\mapsto \exp_{\mathcal{C}(s)}(\textbf{B})$, we construct the Fermi coordinate system $(s,q)$~\cite{klingenberg2013course}. The coordinate $q$ represents the signed geodesic distance along geodesics that emanate orthogonally from the curve $\mathcal{C}(s)$.
For a point $P(s,q)$ on the surface, the metric tensor at $P$ is given by $g_{ab}=(\partial_a \bm{R}(s,q)) \cdot (\partial_b \bm{R}(s,q))$, where $\bm{R}(s,q)$ denotes the location of $P$.

Since the vector $\textbf{B}$ is orthogonal to the tangent vector $\textbf{t}$, this coordinate system is orthogonal. Therefore, the components $g_{sq}=g_{qs}=0$. Additionally, because $\partial_q \bm{R}$ is the tangent vector to a geodesic parametrized by arc length,
\begin{equation}
\left\|\partial_q \bm{R} \right\|=1,
\end{equation}
it follows that $g_{qq}=1$.

Treating $q$ as a small parameter, we we expand $g_{ss}(s,q)$ as a Taylor series:
\begin{equation}
g_{ss}(s,q)=g_{ss}(s,0)+q\partial_q g_{ss}|_{q=0}+\frac{q^2}{2}\partial_q^2 g_{ss}|_{q=0}+O(q^3).
\end{equation}
The zeroth-order term is $g_{ss}(s,0)=\bm{t}\cdot \bm{t}=1$.

For the first-order term,
\begin{equation}
\partial_q g_{ss}=2\partial_q (\partial_s \bm{R})\cdot (\partial_s \bm{R}).
\end{equation}
Here, the derivative should be evaluated in the tangent space $T_{\mathcal{C}(s)}(\mathcal{S})$,
\begin{equation}
\partial_q (\partial_s \bm{R})=\nabla_\textbf{B} \textbf{T},
\end{equation}
where $\textbf{T}=\partial_s \bm{R}$ and $\nabla_\textbf{B} \textbf{T}$ denotes the covariant derivative of $\textbf{T}$ along $\textbf{B}$. Since the surface is torsion-free, $\nabla_\textbf{B} \textbf{T}=\nabla_\textbf{T} \textbf{B}$. Utilizing Eq.~\eqref{dse} for the Darboux frame,
\begin{equation}
\nabla_\textbf{T} \textbf{B}|_{q=0}=(\dot{\bm{B}})_{\tan}=-\kappa_g \textbf{t},
\end{equation}
where the subscript $\tan$ indicates projection onto the tangent plane. Thus,
\begin{equation}
\partial_q g_{ss}|_{q=0}=-2\kappa_g.
\end{equation}
For the second-order term,
\begin{equation}
\partial_q^2 g_{ss}=2[\nabla_{\textbf{B}}(\nabla_{\textbf{B}}\textbf{T})\cdot \textbf{T}+(\nabla_{\textbf{B}}\textbf{T})\cdot(\nabla_{\textbf{B}}\textbf{T})].
\end{equation}
On a two-dimensional surface, the curvature relation
\begin{equation}
R(\textbf{B},\textbf{T})\textbf{B} \cdot \textbf{T}=-K,
\end{equation}
holds, where the Riemann tensor acts as
\begin{equation}
R(\textbf{B},\textbf{T})\textbf{B}=\nabla_{\textbf{B}}(\nabla_{\textbf{T}}\textbf{B})- \nabla_{\textbf{T}}(\nabla_{\textbf{B}}\textbf{B})-\nabla_{[\textbf{B},\textbf{T}]}\textbf{B}.
\end{equation}
Due to commutativity of the coordinate vector fields in two dimensions, $[\textbf{B},\textbf{T}]=0$ and $\nabla_{\textbf{B}}(\nabla_{\textbf{T}}\textbf{B})=\nabla_{\textbf{B}}(\nabla_{\textbf{B}}\textbf{T})$. Furthermore, because the $q$-curve is a geodesic, $\textbf{B}$ is parallel-transported along itself, giving $\nabla_{\textbf{B}}\textbf{B}=0$. Combining these results yields
\begin{equation}
\partial_q^2 g_{ss}|_{q=0}=2[\kappa_g^2-K(s,0)].
\end{equation}
The second-order term can also be derived in an alternative way. Given the metric form $ds^2=dq^2+g_{ss}ds^2$, the Gaussian curvature satisfies $K=-\frac{1}{\sqrt{g}}\partial_q^2\sqrt{g}$. From the relation
$\partial_q^2 g_{ss} = 2 (\partial_q \sqrt{g})^2 + 2 \sqrt{g} \partial_q^2 \sqrt{g}$. Evaluation at $q=0$ confirms $\partial_q^2 g_{ss}|_{q=0}=2[\kappa_g^2-K(s,0)]$, consistent with the prior result.

Note that if we approximate the position of point $P$ using $\bm{R}(s,q)=\bm{r}(s)+q\bm{B}(s)$,
where $|q|$ represents the Euclidean distance from $P$ to $\mathcal{C}$, the first-order term $-2q\kappa_g$ remains valid~\cite{LIANG2025170144}. However, the second-order term deviates from the exact result by $\tau_g^2-K$.

\section{Gauge choice and gauge transformation}\label{appb}
We apply the gauge choice Eq.~\eqref{gau} to the transformation $\bar{\Omega}_q \rightarrow \bar{\Omega}_q+\partial_q(\gamma_b \sigma^b/2)+i[\gamma_b \sigma^b/2,\bar{\Omega}_q]$. The second term on the right side is found to be
\begin{equation}
\partial_q(\gamma_b \sigma^b/2)=- \frac{i}{\sqrt{g}}\epsilon^{bc}\alpha_{qc}\frac{\sigma_b}{2}+\frac{q^2}{2} \partial_q^2(\frac{i}{\sqrt{g}}\epsilon^{bc}\alpha_{qc}\frac{\sigma_b}{2}).
\end{equation}
The first term on the right side cancels $\bar{\Omega}_q$. For the second term, We can find $1/\sqrt{g}=1+O(q)$ and $\alpha_{qc}(s,q)=\sum_{n=0}^{\infty} q^n \frac{\alpha_{ab}^{(n)}(s,0)}{n!}$. The derivative $\partial_q \sigma_b\sim O(1)$ since $[\bar{\Omega}_q,\sigma_b]\sim O(1)$ and the Christoffel symbols $\Gamma_{qa}^{b}\sim O(1)$. In short, the expansion of term $\frac{i}{\sqrt{g}}\epsilon^{bc}\alpha_{qc}\frac{\sigma_b}{2}$ contains no terms of the order $q^{-1}$ and $q^{-2}$. Therefore, we can ensure that
\begin{equation}
\partial_q(\frac{i}{\sqrt{g}}\epsilon^{bc}\alpha_{qc}\frac{\sigma_b}{2})\sim O(1)
\end{equation}
and
\begin{equation}
\partial_q^2(\frac{i}{\sqrt{g}}\epsilon^{bc}\alpha_{qc}\frac{\sigma_b}{2})\sim O(1).
\end{equation}

Similarly, the commutator
\begin{equation}
\begin{aligned}
i[\gamma_b \sigma^b/2,\bar{\Omega}_q]=&i[ -q\bar{\Omega}_q+\frac{1}{2}q^2 \partial_q (\frac{i}{2\sqrt{g}}\epsilon^{bc}\alpha_{qc}\sigma_b),\bar{\Omega}_q] \\
=&\frac{i}{2}q^2 [\partial_q (\frac{i}{2\sqrt{g}}\epsilon^{bc}\alpha_{qc}\sigma_b),\bar{\Omega}_q]\sim O(q^2).
\end{aligned}
\end{equation}
Consequently, after the gauge transformation, $\Omega_q\rightarrow O(q^2)$. 

The second term in Eq.(25) is $-\frac{\hbar^2}{2m}[(\partial_q \bar{\Omega}_q)+q\bar{\Omega}_q^{(1)}\partial_q]$. After the gauge transformation, $(\partial_q \bar{\Omega}_q)\sim O(q)$ and $q\bar{\Omega}_q^{(1)}\partial_q\sim O(q^2)$. As we have rescaled the coordinate $q\rightarrow \epsilon q$, we can confirm that the second term in Eq.(25) is of the order $\epsilon$, which can be neglected in $H_{(0)}$.

The gauge choice can not completely remove the second term in Eq.~\eqref{h0}, however, it can reduce the term to the order of $\epsilon$, which is enough for the derivation of $H_{\text{eff}}$.

\bibliographystyle{apsrev4-1}
\bibliography{ref18}

\end{document}